\documentclass[pra,aps,onecolumn,amsmath,amssymb,showpacs,twocolumn,superscriptaddress,10pt]{revtex4-1}

\usepackage{braket,amsmath,amssymb,amsthm,bm}
\usepackage[caption=false]{subfig}
\usepackage{overpic}
\usepackage{subfig}
\usepackage{xcolor}



\begin{document}

\title{Robust controllability of two-qubit Hamiltonian dynamics}
\author{Ryosuke Sakai}
 %\email{sakai@eve.phys.s.u-tokyo.ac.jp}
 \affiliation{Department of Physics, Graduate School of Science, The University of Tokyo, Tokyo, Japan} 
\author{Akihito Soeda}
%\email{soeda@phys.s.u-tokyo.ac.jp}
 \affiliation{Department of Physics, Graduate School of Science, The University of Tokyo, Tokyo, Japan} 
\author{Mio Murao}
%\email{murao@phys.s.u-tokyo.ac.jp}
\affiliation{Department of Physics, Graduate School of Science, The University of Tokyo, Tokyo, Japan} 
\author{Daniel Burgarth}
%\email{dkb3@aber.ac.uk}
\affiliation{Center for Engineered Quantum Systems, Dept. of Physics \& Astronomy, Macquarie University, 2109 NSW, Australia}

\begin{abstract}

	Physically, quantum gates (unitary gates) for quantum computation are implemented by controlling the Hamiltonian dynamics of quantum systems.  When full descriptions of the Hamiltonians are given, the set of implementable quantum gates is easily characterized by quantum control theory.   In many real systems, however, the Hamiltonians may include unknown parameters due to the difficulty of performing precise measurements or instability of the system. In this paper, we consider the situation that some parameters of the Hamiltonian are unknown, but we still want to perform a {\it robust control} of the Hamiltonian dynamics to implement a quantum gate irrespectively to the unknown parameters.   The existence of the robust control was previously shown for single-qubit systems, and a constructive method was developed for two-qubit systems if a full control of each qubit is available.  We analytically investigate the robust controllability of two-qubit systems, and apply Lie-algebraic approaches to handle the cases where only one of the two qubits is controllable.   We also numerically analyze the robust controllability of the two-qubit systems where the analytical approach is not necessarily applicable and investigate the relationship between the robust controllability of systems with a discrete and continuous unknown parameter.

\end{abstract}
\maketitle

\section{Introduction}

To implement the quantum circuit model of quantum computation \cite{NC:2000} in physical systems, each quantum gate (unitary gate) is generated by Hamiltonian dynamics of the physical systems such as NMR \cite{L:1986, VC:RMP2004}, NV center \cite{DM:PR2013, DBW:NC2014, NAB:PRL2015} and superconducting qubits \cite{FF:N2000, CW:N2008, LMK:PRB2016}.   However, the Hamiltonians of available physical systems may not be the exact generators of the unitary gates.  We can still implement unitary gates by using time-dependent Hamiltonians, if some parts of the Hamiltonian can be changed in time. For a given Hamiltonian and its time-controllable parts, a set of implementable gates can be obtained by quantum control theory \cite{JS:1972, A:PMT2009, DP:IET2010}.   Consider a simple example that a system Hamiltonian is given by $H(t) = u(t)H_0  + v(t)H_1$ where $H_0$ and $H_1$ are fixed time-independent Hamiltonians and $u(t)$ and $v(t)$ are functions representing the time-dependent part that we can control in time.  The functions of $u(t)$ and $v(t)$ are referred to as \textit{control pulses} in quantum control theory and we will assume that they can take both positive and negative values.  The corresponding unitary evolution operator of the Hamiltonian dynamics of $H(t)$ is given by $U(t)=\hat{T}e^{-i \int dt H(t)}$ where $\hat{T}$ represent the time-ordering operator.

There are two key formulas for deriving the set of implementable unitary gates \cite{A:PMT2009, T:1959, S:1976}, i.e., for any bounded operators $A , B$ and $t \in \mathbb{R}_+$,
\\[-5pt]

\textit{(i) Commutator expansion  formula}:
\begin{align}
\begin{split}
	U_{[A, B]}(t) :=&\ e^{-A\sqrt{t}} e^{-B\sqrt{t}} e^{A\sqrt{t}} e^{B\sqrt{t}} \\
	=&\ 1 + [A, B]t + \mathcal{O}(t^{3/2}),
	\label{eq:commutator}
\end{split}
\end{align}

\textit{(ii) Trotter expansion  formula}:
\begin{align}
	U_{A, B}(t) := e^{At} e^{Bt} = 1 + (A + B)t + \mathcal{O}(t^2).
	\label{eq:trotter}
\end{align}
For $H(t) = u(t)H_0  + v(t)H_1$, the commutator expansion guarantees that the Hamiltonian dynamics of $\frac{1}{i}[iH_0, iH_1]$ is simulable by appropriately setting the control pulses $u(t)$ and $v(t)$ as $\lim _{N \to \infty}[U_{[A, B]}(t/N) ]^N = e^{[A,B]t}$.  Here, we implicitly use the condition that the values of $u(t)$ and $v(t)$ can be positive and negative. This formula further implies that dynamics of multiple commutators of $iH_0$ and $iH_1$ such as $\frac{1}{i}[iH_0, [iH_0, iH_1]]$ is also simulable as $e^{[B, A]} = e^{-[A, B]}$, thus the negative times can be simulated.   The Trotter expansion formula verifies that the dynamics of all linear combinations of these multiple commutators of $iH_0$ and $iH_1$ are simulable due to $\lim _{N \to \infty}[U_{A, B}(t/N) ]^N = e^{(A + B)t}$.  Therefore, a set of simulable Hamiltonians has a Lie-algebraic structure, and \textit{any} unitary gates are implementable by appropriately setting $u(t)$ and $v(t)$ if the multiple commutators of the Hamiltonians span all Hermitian operators on the system.	We call such systems to be \textit{fully controllable}.

In fact, one of the control pulses is not necessary for achieving the full control \cite{JS:1972, A:PMT2009, DP:IET2010}.    Consider that the total Hamiltonian is given by $H(t) = H_0 + v(t) H_1$ where $H_0$ is called a drift Hamiltonian constantly applied on the system and we cannot change any part of $H_0$ in time. It is still possible to implement the same set of unitary gates generated by $H(t) = u(t)H_0  + v(t)H_1$ if $H_0$ is a finite dimensional operator,  since we can effectively control the contribution of $H_0$ by setting $v(t)=0$ and simulating the action of its inverse $-H_0$ within an arbitrary error $\varepsilon$ by choosing the evolution time $T' =T_r - t$ where $T_r$ is the recurrence time of $H_0$, i.e., $\Vert I - e^{-i H_0 T_r} \Vert < \varepsilon$. Implementing $H_1$ alone is then a simple consequence of the Trotter expansion formula.

However, if there are \textit{unknown} parameters in a drift Hamiltonian, which happens in real systems \cite{VC:RMP2004, NAB:PRL2015, LMK:PRB2016}, the recurrence time depends on the parameter, and the inverse unitary trick cannot be used.   Nevertheless, it has been shown that there exist \textit{robustly controllable} single-qubit systems with unknown parameters in compact and \textit{continuous} sets \cite{LK:PRA2006, LK:IEEE2009, BCR:CMP2010, BST:EJC2015, BW:SR2015, WB:NC2012}, i.e., we can implement any single-qubit unitary gate irrespective to the unknown parameter by a technique called the polynomial approximation developed by \cite{LK:PRA2006, LK:IEEE2009, BCR:CMP2010, BST:EJC2015} or a systematic search of control pulses for cancelling the unknown parameter \cite{WB:NC2012, BW:SR2015}.

Two-qubit unitary gates are necessary for constructing global unitary operations \cite{Elementarygates:1995, NC:2000} required for quantum computers to outperform classical counterparts.	The robust controllability of two-qubit systems has been explored for the cases where all single-qubit controls are achieved and a general and constructive control method has been proposed \cite{H:PRL2007}.
 
In this paper, we take a more universal method by employing an Lie-algebraic approach to investigate the robust controllability of two-qubit Hamiltonian dynamics even applicable for the cases where the control pulse is available on the Hamiltonian of only one of the two-qubits.  We also consider the robust controllability of the two-qubit systems where the values of the unknown parameters are given as finite sets, or a \textit{discretized} subset of a given continuous set.	The robust controllability of the systems with unknown parameters given in finite (discretized) sets has been well studied \cite{BST:EJC2015, TVLR:JPA2004, DHR:MCS2016}, but the differences between continuous and discretized cases has not been clarified, i.e., whether or not the robust controllability of the systems with unknown parameters in all discretized subsets of a given continuous set implies the robust controllability of the continuous one.  We investigate this problem by combining analytical and numerical approaches.

This paper is organized as follows.    We briefly review the proofs of the robust controllability of given single-qubit systems in Sec.~\ref{sec:RC1}, and show the robust controllability of two-qubit systems in Sec.~\ref{sec:RC2}. We also give systems whose robust controllability is unclear by our analytical approach in this section.	In Sec.~\ref{sec:D}, we introduce another technique, discretization, to numerically investigate their robust controllability by using the QuTip control package \cite{JNN:QuTip2012, JNN:QuTip2013}, and provide an example which does not seem robust controllable in Sec.~\ref{sec:NR}.  We also discuss the relation between robust controllability for unknown parameters in continuous and discretized sets in Sec.~\ref{sec:NR}.	Finally, we present the summary and discussions in Sec.~\ref{sec:Conclusion}.

\section{Review of the robust control for single-qubit systems}
\label{sec:RC1}

The total Hamiltonian of the system is given by $H(t) = H_d (\omega) + v(t) H_c$ where the drift Hamiltonian $H_d(\omega )$ contains an unknown parameter $\omega$ and $H_c$ is the part of the Hamiltonian called the control Hamiltonian associated with the control pulse $v(t)$. We allow $v(t)$ to consist of arbitrary piecewise constant elements and the bang-bang style Delta function pulses \cite{VL:PRA1998,VKL:PRL1999,VLK:PRL1999}.	It turns out to be important to separate the drift Hamiltonian from the controllable part by the control pulse for the case with unknown parameters. We also assume $\omega$ is in $[\omega _0, \omega _1] \subset \mathbb{R}$ and denote the set of Hamiltonians which can be simulated by $H(t)$ as $\mathbb{L}_{H(t)}$.

The first example \cite{LK:PRA2006, LK:IEEE2009, WB:NC2012} of the robust control of the single-qubit system is presented in the system whose Hamiltonian is $H_d(\omega) = \omega X$ and $H_c = Z$, where $X, Y$ and $Z$ are Pauli operators.  By applying Delta function pulses we can effectively apply the unitary gate $Z$ and thereby invert $H_d (\omega)$, i.e., $-H_d(\omega) \in \mathbb{L}_{H(t)}$ because of $ZXZ = -X$, and $\alpha H_d(\omega) + \beta H_c \in \mathbb{L}_{H(t)}$ by the Trotter expansion formula.	According to the commutator expansion formula, $\omega Y = [iH_d(\omega), iH_c]/(2i)$ is in $\mathbb{L}_{H(t)}$ and $\omega ^2 H_c = [i \omega Y, iH_d(\omega)]/(2i) \in \mathbb{L}_{H(t)}$.	By induction, the dynamics of $\omega ^{2n + 1}X, \omega ^{2n + 1}Y$ are implementable for $n = 0, 1, 2, \cdots$. Taking linear combinations with weighted coefficients and using the Trotter expansion formula, we can see $\omega f_1 (\omega ^2) X + \omega f_2 (\omega ^2) Y  \in \mathbb{L}_{H(t)}$ for any polynomial functions $f_i$'s.  $f_i(\omega)$ can be approximated so that $\omega f_1(\omega ^2) = \theta _1$ and $\omega f_2 (\omega ^2) = \theta _2$ within an arbitrary small error for any $\omega \in [\omega _0, \omega _1]$ and constant $\theta _1, \theta _2 \in \mathbb{R}$ if $\omega _0 \omega_1 > 0$ because $\omega f_i(\omega^2)$ is odd. Having any rotation around the $X$ and $Y$ axes at hand, we can robustly perform any quantum gate on this system when $\omega_0 \omega_1 > 0$, and this technique is called the \textit{polynomial approximation} \cite{LK:PRA2006, LK:IEEE2009, BCR:CMP2010}.

The second example is $H_d(\omega) = X + \omega Y$ and $H_c = Z$. This has the same controllability of $u(t)H_d(\omega) + v(t)H_c$ since $-H_d(\omega) = ZH_d(\omega)Z$. By a similar procedure of the first example, we can see 
\begin{align*}
   &\ f_1 (\omega^2)H_d(\omega) + f_2 (\omega^2)H_c + f_3(\omega^2)H_{2}(\omega) \nonumber \\
   =&\ (f_1 (\omega ^2) - \omega f_3(\omega ^2))X  \nonumber \\
   &\hspace{40pt} + (f_3(\omega ^2) + \omega f_1(\omega ^2))Y + f_2 (\omega ^2) Z
	%\label{eq:ex3_ache}
\end{align*}
where $H_2(\omega) := Y - \omega X$  is simulable \cite{BST:EJC2015} for arbitrary polynomial functions $f_i(\omega ^2)$.  Assume there are robustly controllable functions $f_1, f_3$ satisfying $f_1(\omega ^2) - \omega f_3 (\omega ^2) = \theta _1$ and $f_3 (\omega ^2) + \omega f_1 (\omega ^2) = \theta _2$ for any $\theta _1, \theta _2 \in \mathbb{R}$. Then
\begin{align}
	f_1 (\omega ^2) = \frac{\theta _1 + \omega \theta _2}{1 + \omega ^2}, \ \ f_3 (\omega ^2) = \frac{\theta _2 - \omega \theta_1}{1 + \omega ^2}
	\label{eq:f1f3}
\end{align}
are required, and if $\omega _0 \omega _1 > 0$, either $\omega = \sqrt{\omega ^2}$ or $-\omega = \sqrt{\omega ^2}$ is satisfied for all $\omega \in [\omega _0, \omega _1]$, thus the right hand sides of Eqs.~(\ref{eq:f1f3}) are described by polynomials of $\omega ^2$ within an arbitrary accuracy, and this system is robustly controllable as long as $\omega _0 \omega _1 > 0$.

\section{Robustly controllability of two-qubit systems}
\label{sec:RC2}

We show that there exist robustly controllable two-qubit systems for a compact and continuous unknown parameter with the polynomial approximation in Sec.~\ref{ssec:PRC2}.   In Sec.~\ref{ssec:URC2}, we will give systems for which we cannot show the robust controllability by the polynomial approximation.   Whether these systems are robustly controllable or not is an open problem.

\subsection{Proofs of robust controllability by the polynomial approximation}
\label{ssec:PRC2}

We show four robustly controllable two-qubit systems in this subsection.   The total Hamiltonian of each system has either one or two control Hamiltonians, e.g. $H(t) = H_d (\omega) + v(t) H_c$ or $H(t) = H_d (\omega) + u(t)H_{c_1} + v(t)H_{c_2}$.	The first two systems (System A and B) have one control Hamiltonian, and the unknown parameter $\omega$ of System A and B is on a local Hamiltonian of one of the two qubits and on an interaction Hamiltonian between two qubits, respectively.   The other two systems (System C and D) have two control Hamiltonians, and their unknown parameter $\omega$ is the coupling strength of the two-qubit Heisenberg interaction.   Also, System D has a second unknown parameter $\nu$ corresponding to an additional local field.

\textit{System A -- $H_d(\omega ) = \omega X \otimes I + X \otimes X + Y \otimes Y + Z \otimes Z$ and $H_c = Z \otimes I$.	(One unknown parameter $\omega$ on a local Hamiltonian and one control pulse)}:	
This system can simulate the Hamiltonians dynamics of the following Hamiltonians; 
\begin{align}
	H_d (\omega) \ &\mathrm{and}\  \pm H_c, \label{eq:2q-ex1_1} \\
	- H_d (\omega) + 2Z \otimes Z \ &\mathrm{and} \ \pm Z \otimes Z, \label{eq:2q-ex1_2}\\
	- H_d (\omega)\ &\mathrm{and} \ \pm \omega X \otimes I, \label{eq:2q-ex1_3}\\
	\pm (X \otimes X +\ &Y \otimes Y + Z \otimes Z) \label{eq:2q-ex1_4}.
\end{align}
The Hamiltonians in (\ref{eq:2q-ex1_1}) are simulable by the Delta function technique for $v(t)$.   From the observation of $(Z \otimes I) [H_d (\omega)] (Z \otimes I) = -H_d(\omega) + 2 Z \otimes Z$, we obtain Hamiltonians in (\ref{eq:2q-ex1_2}) by the Trotter formula and the finding recurrence time of $Z \otimes Z$.  Hence, the Hamiltonians in (\ref{eq:2q-ex1_3}) are simulable by the Trotter expansion formula and the commutator expansion formula since $\omega X \otimes I \propto [Z \otimes Z, [H_d(\omega), Z \otimes Z]]$, and finally Hamiltonians given by (\ref{eq:2q-ex1_4}) are obtained by the Trotter expansion formula.

We showed in Sec.~\ref{sec:RC1} that adjusting $\pm \omega X$ and $\pm Z$ is sufficient to robustly control single-qubit dynamics for $\omega \in [\omega_0, \omega_1]$ if $\omega _0 \omega_1 > 0$ is satisfied.  Full control for one of the qubits and the two-qubit Heisenberg interaction achieve the full controllability of the two-qubit system, thus we can implement any unitary gate in SU(4) on the system as long as $\omega_0 \omega _1 > 0$.  

This procedure also works for the system with $H_d(\omega) = X \otimes I + \omega Y \otimes I + X \otimes X + Y \otimes Y + Z \otimes Z$ and $H_c = Z \otimes I$.  That is, we can obtain the same simulable Hamiltonians given by (\ref{eq:2q-ex1_1}-\ref{eq:2q-ex1_4}) except the right Hamiltonian of (\ref{eq:2q-ex1_3}), i.e., $\omega X \otimes I \to X \otimes I + \omega Y \otimes I$.    Adjusting $X \otimes I + \omega Y \otimes I$ and $Z \otimes I$ is sufficient to robustly control single-qubit dynamics, therefore this system is also robustly controllable.

\textit{System B -- $H_d (\omega) = X \otimes I + I \otimes X + \omega (X \otimes X + Y \otimes Y)$ and $H_c = Z \otimes I$.	(One unknown parameter $\omega$ on an interaction Hamiltonian and one control pulse)}:    The simulable Hamiltonians of this system are
\begin{align}
   H_d (\omega)\ &\mathrm{and} \ \pm Z \otimes I, \label{eq:2q-exb_1} \\
   - H_d (\omega) + 2I \otimes X\ &\mathrm{and}\ \pm I \otimes X, \label{eq:2q-exb_2} \\
   - H_d (\omega)\ &\mathrm{and} \ \pm \omega Y \otimes Y. \label{eq:2q-exb_3}
	% \pm \omega ^2 (ZI - IZ)&, \ \pm \omega ^{2n + 1}(XX + YY), 
\end{align}
The procedure of obtaining Hamiltonians in (\ref{eq:2q-exb_1}-\ref{eq:2q-exb_3}) are the same as (\ref{eq:2q-ex1_1}-\ref{eq:2q-ex1_3}) due to $\omega Y \otimes Y \propto [I \otimes X, [I \otimes X, H_d (\omega )]$.  By the commutator expansion formula, $\omega ^2 I \otimes Y \propto [I \otimes X, [H_d (\omega ), [Z \otimes I, H_d (\omega)]]], \ \omega ^2 I \otimes Z \propto [I \otimes X, \omega ^2 I \otimes Y]$ and $\omega ^4 I \otimes X \propto [\omega ^2 I \otimes Y, \omega ^2 I \otimes Z]$ are simulable, thus $\omega ^{4n + 2} I \otimes Y \propto [\omega ^4 I \otimes X, \omega ^{4n - 2} I \otimes Z]$ and $\omega ^{4n + 2} I \otimes Z \propto [\omega ^4 I \otimes X, \omega ^{4n - 2} I \otimes Y]$ are inductively simulable for any $n = 1, 2, \cdots$, and $I \otimes Y$ is robustly simulable by the polynomial approximation if $[\omega _0, \omega_1]$ does not include zero.	Now we obtain $\omega X \otimes X \propto [Z \otimes I, [I \otimes Z, \omega Y \otimes Y]]$ and thus $X \otimes I + I \otimes X$ by the Trotter expansion formula.	$X \otimes X$ and $Y \otimes Y$ are robustly simulable via multiple commutators of $\omega ^2 I \otimes Z$ if $\omega _0 \omega _1 > 0$.	System B is robustly controllable because controlling $X \otimes I + I \otimes X, \ I \otimes Y, \ X \otimes X$ and $Y \otimes Y$ is sufficient to implement any unitary gates in SU(4).

\textit{System C -- $H_d (\omega) = \omega (X \otimes X + Y \otimes Y + Z \otimes Z)$, $H_{c_1} = X \otimes I$ and $H_{c_2} = Z \otimes I$.	(One unknown parameter $\omega$ on an interaction Hamiltonian and two control pulses)}:    The simulable Hamiltonians of this system are
\begin{align}
   H_d (\omega)\ \mathrm{and}\ &\pm X \otimes I \ \mathrm{and} \ \pm Z \otimes I, \label{eq:2q-exc_1} \\
   \begin{split}
      - H_d (\omega) + 2\omega X \otimes X\ & \mathrm{and} \ - H_d (\omega) + 2\omega Y \otimes Y \\
      \mathrm{and} \ -& H_d (\omega) + 2\omega Z \otimes Z,
   \end{split} \label{eq:2q-exc_2} \\
   \pm \omega X \otimes X\ \mathrm{and} \ &\pm \omega Y \otimes Y \ \mathrm{and} \pm \omega Z \otimes Z. \label{eq:2q-exc_3}
\end{align}
Hamiltonians in (\ref{eq:2q-exc_2}) are obtained by applying strong local fields, i.e., $\sigma _i H_d (\omega) \sigma _i^\dagger$ for $i = 1,2,3$, where $\sigma _1 = X \otimes I,\ \sigma _2 = (Z \otimes I)(X \otimes I)$ and $\sigma _3 = Z \otimes I$, respectively.  By the Trotter expansion formula and the Delta function technique, we can simulate Hamiltonians in (\ref{eq:2q-exc_3}) and $\pm \omega ^{2n + 1} X \otimes X,\ \pm \omega ^{2n + 1} Y \otimes Y$ and $\pm \omega ^{2n + 1} X \otimes X$ for $n = 0, 1, 2, \cdots$ since $\omega ^{2n + 1} X \otimes X \propto [Z \otimes I, [\omega X \otimes X, [\omega X \otimes X, [\omega ^{2n - 1}X \otimes X, Z \otimes I]]]]$ holds and so on.  Thus we obtain $X \otimes X,\ Y \otimes Y$ and $Z \otimes Z$ dynamics by the polynomial approximation if $\omega \in [\omega _0, \omega _1]$ and $\omega _0 \omega _1 > 0$, and the robust controllability of System C is shown by using the full controllability of each single qubit and the Heisenberg interaction.

\textit{System D -- $H_d (\nu, \omega) = \nu X \otimes I + \omega (X \otimes X + Y \otimes Y + Z \otimes Z)$, $H_{c_1} = X \otimes I$ and $H_{c_2} = Z \otimes I$. (Two unknown parameters $\nu, \omega$ and two control pulses)}:    The simulable Hamiltonians of System D are almost same as the Hamiltonians given by (\ref{eq:2q-exc_1}-\ref{eq:2q-exc_3}) except $\omega X \otimes X \to \nu X \otimes I + \omega X \otimes X$.	We can robustly simulate $Y \otimes Y$ and $Z \otimes Z$ similarly to System C, and $\pm \nu X \otimes I \propto [Y \otimes Y, [\nu X \otimes I + \omega X \otimes X, Y \otimes Y]]$ are obtained, i.e., canceling $\omega X \otimes I$ is possible.	Thus System D is robustly controllable for $\omega \in [\omega _0, \omega _1]\ (\omega _0 \omega _1 > 0)$ and any $\nu \in \mathbb{R}$ in principle.

	In Systems A to D, we introduced three techniques to show robust controllability, obtaining a set of Hamiltonians whose inverse dynamics are simulable, showing simulable Hamiltonians generated by the set via Lie-algebraic approach and the polynomial approximation.	However, there are the cases where we cannot obtain a large enough set of invertible Hamiltonians to algebraically show robust controllability as presented in the next subsection.

\subsection{Systems whose robust controllability is unclear}
\label{ssec:URC2}

We show examples of the systems whose robust controllability is unclear via the polynomial approximation in this subsection.

\textit{System E -- $H_d (\omega) = X \otimes I + \omega (X \otimes X + Y \otimes Y + Z \otimes Z)$ and $H_{c} = Z \otimes I$. (One unknown parameter $\omega$ on an interaction Hamiltonian and one control pulse)}:	  Simulable Hamiltonians are the following ones obtained by the same procedure of System A:
\begin{align}
	H_d (\omega)\ &\mathrm{and} \pm Z \otimes I, \label{eq:2q-ex3-1} \\
	-H_d (\omega) + 2 \omega Z \otimes Z \ &\mathrm{and} \ \omega Z \otimes Z. \label{eq:2q-ex3-2}
\end{align}
However, the simulability of $-\omega Z \otimes Z$ is unclear since the recurrence time of $\omega Z \otimes Z$ depends on the unknown parameter $\omega$.	Also strong fields do not work because $\omega Z \otimes Z$ is commuting with $H_c$.	Hence, we cannot apply the procedure for obtaining (\ref{eq:2q-ex1_2}, \ref{eq:2q-ex1_3}), and the Hamiltonians (\ref{eq:2q-ex3-1}, \ref{eq:2q-ex3-2}) are \textit{not enough} to prove the robust controllability of System E.	Whether this system is robustly controllable or not is unclear from the polynomial approximation, as there may be less direct ways involving non-algebraic evolutions leading to robust elements.

This kind of problem happens even in a single-qubit system such as $H_d (\omega) = X + \omega Z$ and $H_c = Z$.   By applying strong fields, we can simulate $Z(-H_d (\omega)) Z = -H_d(\omega) + 2\omega Z$, and $\omega Z$ by the Trotter expansion formula.  However, simulability of $- \omega Z$ is unclear because $\omega Z$ and $H_c$ commute, and thus the robust controllability of this system is also unclear with our Lie-algebraic approach.  Note that, however, it does \textit{not} necessarily imply that these systems are not robustly controllable.	 

In the following section, we will numerically investigate the robust controllability by using a method called \textit{discretization} \cite{BST:EJC2015, TVLR:JPA2004, DHR:MCS2016} for a given region of the unknown $\omega$ to provide a robust control pulse and investigate the robust controllability of System E for the region.	In addition, System E is a good candidate to see the difference between the robust controllability for continuous and discretized unknown parameters as mentioned in Sec.~\ref{sec:controllability_of_E}.

\section{Discretization of the unknown parameter}
\label{sec:D}

In this section, we introduce another method to seek robust controllability, \textit{discretization}. The idea is to make sure that the controls are robust on equally spaced points in the interval and hope that the robust controllability is kept in other points between them. This may seem to be a natural strategy, but with increasing number of points the control time also increases, and thus it is not clear if this strategy works toward the continuous limit.

Specifically, consider $\omega$ in a finite set $\Omega _N = \{ \omega ^{(1)}, \omega ^{(2)}, \cdots , \omega ^{(N)} \} \subset [\omega _0, \omega _1]$.  In this case, we only need to guarantee the robustness for $N$ different configurations of the systems whose Hamiltonians are given by $H_d (\omega ^{(n)}) + v(t)H_c$ for $n=1,2,\cdots , N$.	Our goal is to implement any target unitary gate $U$ on each configuration of the systems in the same time.	The robust controllability for $\omega \in \Omega _N$ is guaranteed by quantum control theory, i.e., the system can be described by a larger dimensional system with a fully \textit{known} system by defining another total Hamiltonian $\bar{H}(t) = \bar{H}_d(\Omega _N) + v(t)\bar{H}_c$ where 
\begin{align*}
	\bar{H}_d (\Omega _N) &= \oplus _{n=1}^N H_d(\omega ^{(n)}), \\
	&=\left[
	\begin{array}{cccc}
		H_d(\omega ^{(1)}) & & & \\
		& H_d (\omega ^{(2)}) & & \\
		& & \ddots & \\
		& & & H_d (\omega ^{(N)})
	\end{array}
	\right], \\
	\bar{H}_c &= \oplus _{n=1}^N H_c.
\end{align*}
are given in block-diagonal forms.	The recurrence time of the extended fully known system $\bar{H}_d(\Omega_N)$ can be derived in principle, thus this system has the same controllability of $u(t) \bar{H}_d(\Omega_N) + v(t)\bar{H}_c$.	Now, the problem to be solved becomes the number of linearly independent Hamiltonians generated by the multiple commutators between $\bar{H}_d(\Omega_N)$ and $\bar{H}_c$.	The maximal number is $N (d^2 - 1)$ since $\bar{H}(t)$ are in the block-diagonal form where $d$ is the dimension of the individual systems.   To achieve the maximal number, there is a useful lemma for controllability of such a block-diagonalized Hamiltonian system \cite{BST:EJC2015, TVLR:JPA2004, DHR:MCS2016}:	The system with $H_\omega (t) := H_d (\omega) + v(t) H_c$ is robustly controllable for $\omega \in \Omega _N$ within an arbitrary small error if and only if 
\begin{itemize}
\item[\textit{(1)}] \textit{The system with Hamiltonian $H_\omega (t)$ is fully controllable for each $\omega \in \Omega _N$.}	
\item[\textit{(2)}] \textit{All Hamiltonians $H_{\omega \in \Omega _N} (t)$ are not mutually unitarily equivalent.}
\end{itemize}
Note that an operator $A$ is unitarily equivalent to another operator $B$ via a unitary operator $X$ if the relation $A = X B X^\dagger$ holds.

The first condition guarantees to implement an arbitrary quantum gate on the individual systems.	The second condition is required to perform quantum gates on each of the systems independently.	Note that the first condition is not necessary if full controllability is not required, and satisfying the second condition implies not only robust but also \emph{ensemble} controllability, i.e., we can implement $\bar{U} = \oplus _{i=1}^N U_i$ for any $U_i$'s in SU($d$).  A special case of ensemble control where all $U_i$ for all $i$ are identical to a given target gate $U$, namely $U_i = U$ for all $i$ corresponds to robust control.

From the second condition, we can see the reason why $\omega _0 \omega _1 > 0$ is required in the case of $H_d (\omega) = \omega X$ and $H_c = Z$.	It is trivial for $\omega = 0$, thus we assume that $\omega _0 \omega _1 < 0$, i.e., $\omega _0 < 0 < \omega _1$, then there exists $\xi > 0$ such that $\pm \xi \in [\omega _0, \omega _1]$.	$H_\xi (t)$ is unitarily equivalent to $H_{-\xi} (t)$ via $Z$ and the second condition is violated.

Discretizing the unknown parameter is a useful technique for numerically searching an appropriate $v(t)$, as the search can be reduced for finding a robust control pulse $v(t)$ to implement $\bar{U} = \oplus _{i=1}^N U$ by the fully known block-diagonal Hamiltonian $\bar{H}(t)$ for a target gate $U$.  The Gradient Ascent Pulse Engineering (GRAPE) algorithm \cite{K:GRAPE} is a well known method to solve this kind of problems although there are many other methods to search control pulses \cite{KGB:PRA2013, BM:PRA2013, DRSG:PRL2013, DSG:PRA2017, CDL:PRA2014, DWC:SR2016, CDQ:IEEE2017}.	We use the QuTip control package \cite{JNN:QuTip2012, JNN:QuTip2013} to find the robust control pulse $v(t)$ by the discretization approach in Sec.~\ref{sec:NR}. 

By using numerical searches with the discretization, we can estimate the control time $T_\varepsilon (N)$ to achieve $\varepsilon$ error for all $N$ points.  If the scaling of $T_\varepsilon (N)$ is less than $\mathcal{O}(N)$, then we can see the worst error between $N$ points becomes smaller with increasing $N$. To see this, we use the inequality \cite{ARBR:NJP2017}
\begin{align}
   \Vert U_{\omega _a}(t) - U_{\omega _b}(t) \Vert \leq t \Vert H_d (\omega _a) - H_d (\omega _b) \Vert
   \label{eq:worsterror}
\end{align}
where $U_\omega (t) = \hat{T}e^{-i \int _0^t d\tau H_\omega (\tau )}$.  For any $\omega _a \in [\omega _0, \omega _1]$, there exists $\omega _b \in \Omega _N$ such that $\Vert H_d (\omega _a) - H_d (\omega _b) \Vert \propto |\omega _a - \omega _b | \leq 1/2(N-1)$, and $\Vert H_d (\omega _a) - H_d (\omega _b) \Vert T_\varepsilon (N)$ decreases with $N$ if $T_\varepsilon (N) < \mathcal{O}(N)$. Thus, the robust controllability for a continuous unknown parameter can be clear with respect to a given allowed error by estimating $T_\varepsilon (N)$.   Although the researches in \cite{WB:NC2012, BW:SR2015} show methods to obtain robust control pulses for several single-qubit systems, two-qubit and more general single qubit cases are still unclear.    Thus, we use the discretizing approach to see the robust contollability.

\section{Numerical results of robust control}
\label{sec:NR}

The analytical result provides the existence of a control pulse $v(t)$ approximating any unitary gate in \rm{SU}(4) in arbitrary accuracy for a compact and positive (or negative) continuous parameters, but it does not provide the construction of $v(t)$.	We investigate whether the pulse sequences numerically obtained by the discretization approach can be also applicable to achieve the robust control for the continuous range of the corresponding unknown parameter \cite{KGB:PRA2013, BM:PRA2013, DRSG:PRL2013, DSG:PRA2017, CDL:PRA2014, DWC:SR2016, CDQ:IEEE2017}, and whether the applicability depends on the types of Hamiltonians whose robust controllability for a continuous unknown parameter is analytically shown (System A) or not (System E).

\subsection{Numerically searching robust control pulses by discretization}

We show the discretization trick helps to find the robust control pulse for a \textit{continuous} unknown parameter $\omega \in [\omega _0, \omega _1]$ by an example of System A where the robust controllability is analytically shown.  For the numerical search of $v(t)$, we first choose $\Omega _{11} = \{ 1.0, 1.1, \cdots, 2.0 \} \subset [1,2]$, i.e., $\omega _0 = 1$ and $\omega _1 = 2$, and a CNOT gate as the target unitary gate.  We obtain $v(t)$ by using the QuTip control package \cite{JNN:QuTip2012, JNN:QuTip2013}.   The accuracy of robust control is evaluated by the error $\epsilon (\omega)$ of the dynamics $U_\omega (T)$ generated by $H_d(\omega) + v(t) H_c$ with a control time $T$ against the target gate operation $U_\mathrm{targ}$ given by  
\begin{align}
	\epsilon(\omega) = 1 - (1/d) \mathrm{Tr}(U_\mathrm{targ}^\dagger U_\omega (T))
	\label{eq:error}
\end{align}
where $d$ is the dimension of the system and $d=4$ is selected in the numerical analysis.

\begin{figure}[tb]
\centering
\begin{overpic}[width=8.6cm]{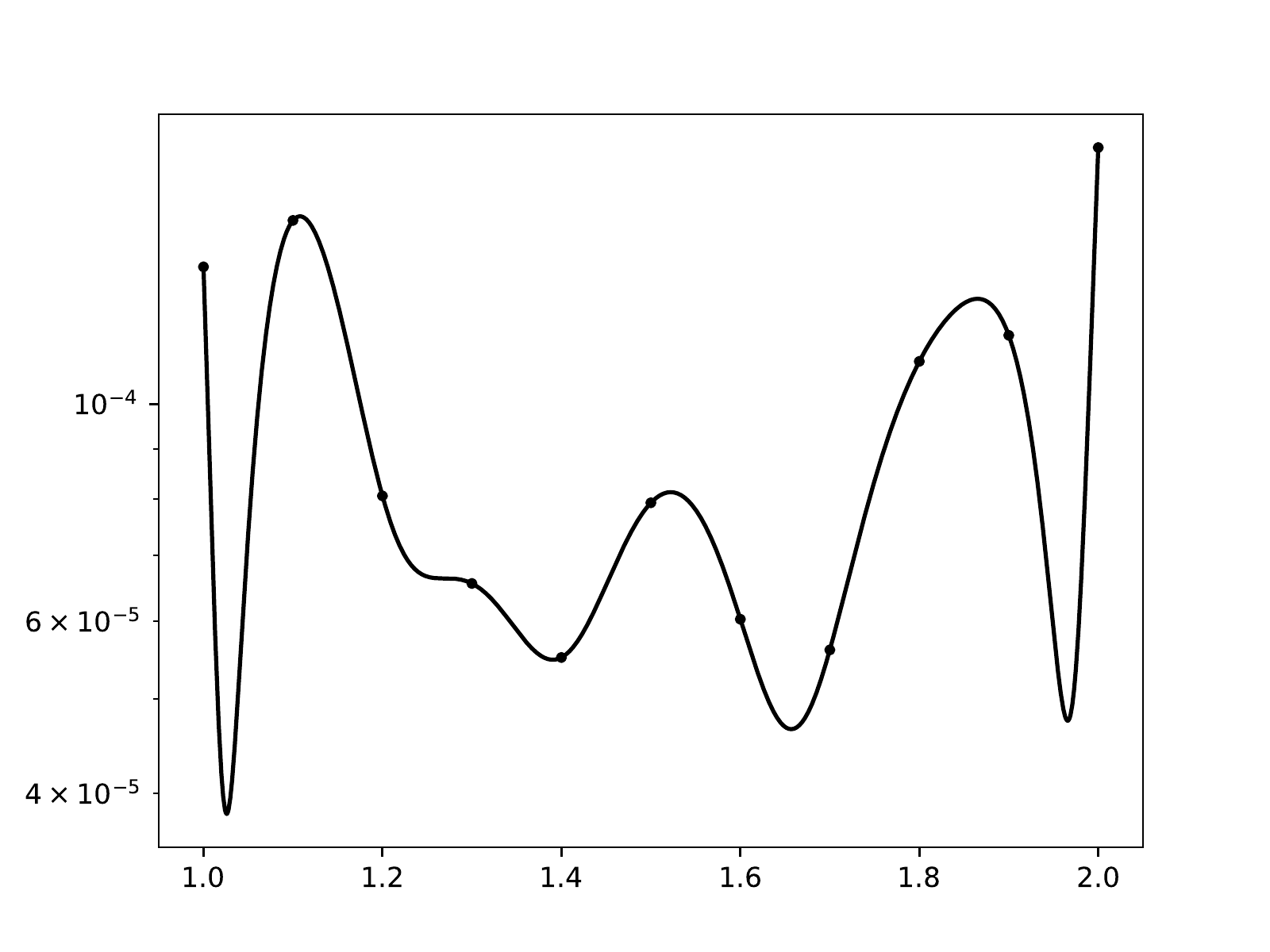}
	\put(50,1){$\omega$}
	\put(5,68){$\epsilon (\omega)$}
\end{overpic}
\caption{The robust control for $\omega \in [1, 2]$ to implement a CNOT gate by the dynamics of $H(t) = H_d(\omega ) + v(t)H_c$ for $H_d (\omega) = \omega X \otimes I + X \otimes X + Y \otimes Y + Z \otimes Z$ and $H_c = Z \otimes I$ (System A), where $v(t)$ is numerically obtained by the discretization approach for $\Omega _{11} = \{ 1.0, 1.1, \cdots, 2.0 \}$.  The black line represents errors (a vertical axis) of the dynamics with $\omega \in [1,2]$ (a horizontal axis) from the ideal CNOT gate evaluated by Eq.~(\ref{eq:error}) where $d = 4$ and $T = 32$.	The black dots represent errors for the points in $\Omega _{11}$.}
\label{fig:omegaXI}
\centering
% \begin{overpic}[scale=0.35]{./figs/X_128ts_50p.pdf}
\begin{overpic}[width=8.6cm]{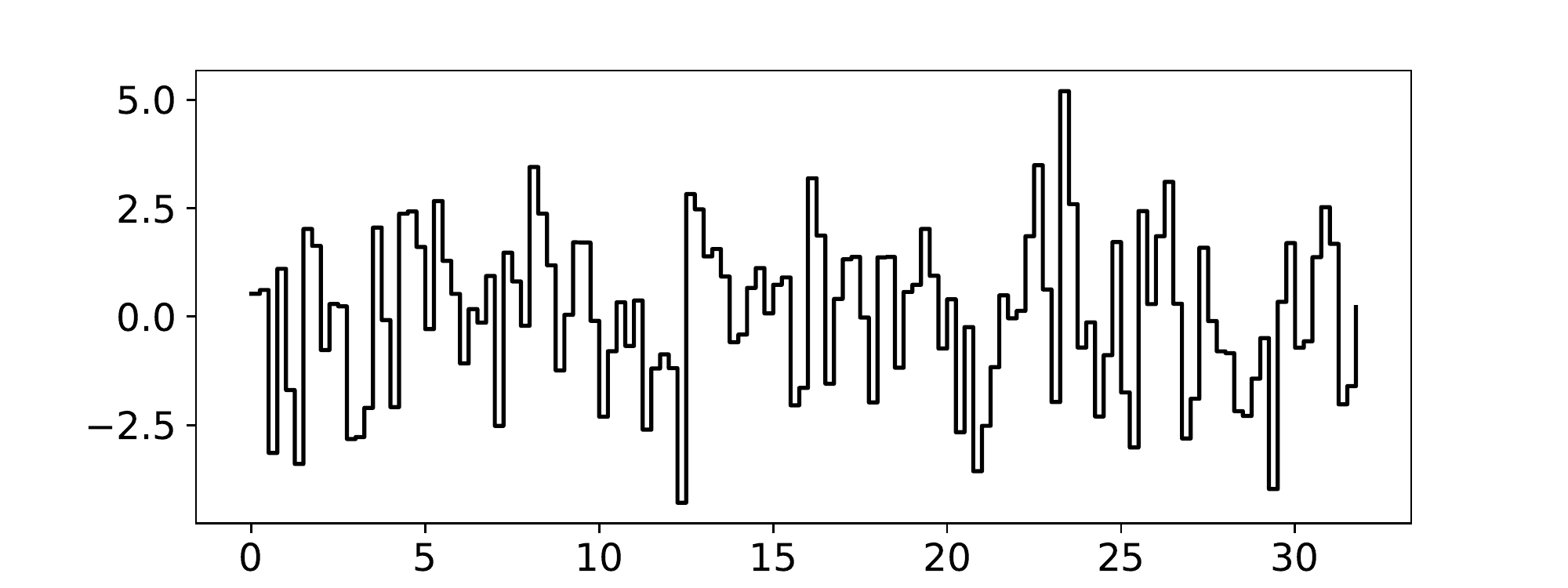}
	\put(52,-3){\small $t$}
\end{overpic}
\caption{The control pulse $v(t)$ for the robust control realizing the error spectrum shown in Fig.~\ref{fig:omegaXI}.  The total time of control $T$ is 32.	We divide the total time by 128, and set $v(t)$ be a constant in $t \in [(n - 1)T/128, nT/128]$ for any $n = 1, 2, \cdots 128$.}
\label{fig:omegaXI_pulse}
\end{figure}

Fig.~\ref{fig:omegaXI} represents the error spectrum $\varepsilon (\omega)$ over $\omega \in [1, 2]$ (the black line), and the black dots represents the errors $\epsilon (\omega)$ for the discrete points in $\Omega _{11}$, where the control time is $T=32$.   The error spectrum is kept around $\mathcal{O}(10^{-4})$ and the robust control is achieved in $\omega \in [1, 2]$ despite the fact that $v(t)$ is obtained by the algorithm guaranteeing the accuracy only for $\omega \in \Omega _{11}$.  Fig.~\ref{fig:omegaXI_pulse} represents the control pulse $v(t)$ generating the robust control of Fig.~\ref{fig:omegaXI}.	Note that the Delta function-like large amplitude control pulses are necessary to show the robust controllability for the continuous unknown parameter analytically by the method presented in Sec.~\ref{sec:RC1} and \ref{sec:RC2}, but such pulses do not appear in Fig.~\ref{fig:omegaXI_pulse}.

\subsection{Robust controllability of System E}
\label{sec:controllability_of_E}

\begin{figure}[tb]
\centering
\begin{overpic}[width=8.6cm]{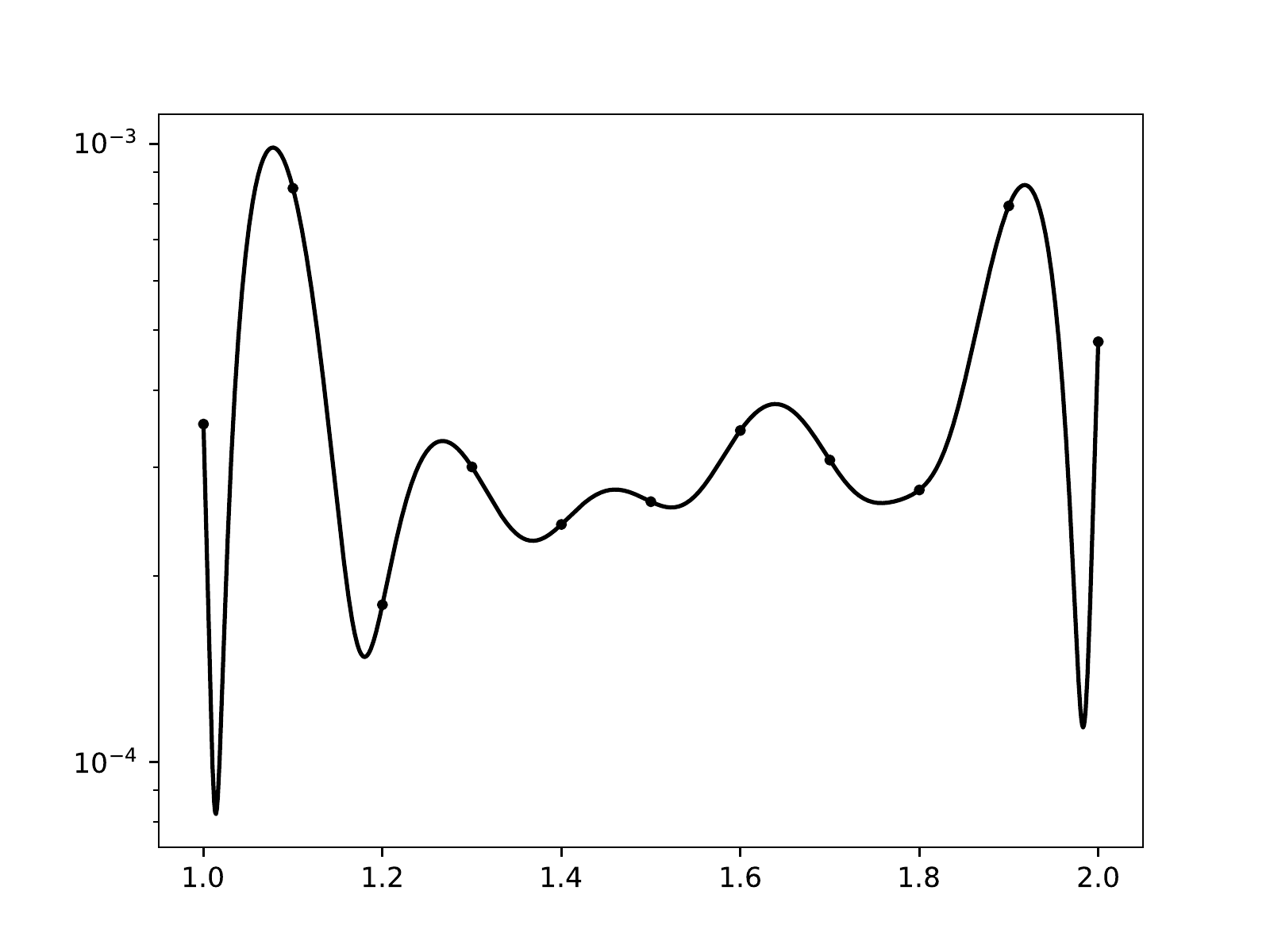}
	\put(50,1){$\omega$}
	\put(5,68){$\epsilon (\omega)$}
\end{overpic}
\caption{The robust control for $\omega \in [1, 2]$ to implement a CNOT gate by the dynamics of $H(t) = H_d(\omega ) + v(t)H_c$ for $H_d (\omega) = X \otimes I + \omega (X \otimes X + Y \otimes Y + Z \otimes Z)$ and $H_c = Z \otimes I$ (System E), where $v(t)$ is numerically obtained by the discretization approach for $\Omega _{11} = \{ 1.0, 1.1, \cdots, 2.0 \}$.  The black line represents errors (a vertical axis) of the dynamics with $\omega \in [1,2]$ (a horizontal axis) from the ideal CNOT gate evaluated by Eq.~(\ref{eq:error}) where $d = 4$ and $T = 32$.	The black dots represent errors for the points in $\Omega _{11}$.}
\label{fig:omegaHei1}
\end{figure}

We investigate the robust controllability of System E by numerical approaches where the robust controllability is unclear within our analytical approach.   System E with $H_\omega (t)$ is fully controllable for each $\omega \in [1, 2]$ and all $H_\omega (t)$ and $H_{\omega'} (t)$ with $\omega \neq \omega ' \in [1, 2]$ are \textit{not} mutually unitarily equivalent, thus there exists a robust control pulse $v(t)$ for $\omega \in \Omega _{11}$ in an arbitrary small error according to the lemma presented in the previous section.   In this sense, investigating the robust controllability of System E helps to understand the relations between the robust controllability for continuous and discretized unknown parameters.	To search the robust control pulse for a continuous unknown parameter $\omega \in [1, 2]$, we choose $\Omega _{11} := \{ 1.0, 1.1, \cdots, 2.0 \} \subset [1, 2]$, the CNOT gate as the target gate and $T=32$.   We obtain Fig.~\ref{fig:omegaHei1} whose black dots represents the errors of the dynamics generated by $H_d(\omega) + v(t) H_c$ with a duration time $T = 32$ for the discrete points in $\Omega _{11}$, and the black line represents $\epsilon (\omega)$ over $\omega \in [1,2]$ by using the same control pulse $v(t)$ for the cases of both discretized and continuous unknown parameters.  The error spectrum is kept around $\mathcal{O}(10^{-4})$ over $\omega \in [1, 2]$.

	The question now is whether we can obtain the robust control pulse for an arbitrary small allowed error.   To see this, we numerically estimate the minimum control time $T_\varepsilon (N)$ for given $N$ and $\varepsilon$, where $N$ is the number of discretization such as $\Omega _N := \{1 + \frac{n}{N - 1} \mid n = 0, 1, \cdots , N -1 \} \in [1, 2]$, and $\varepsilon$ is an allowed error for each configuration.  We set the target gate to the CNOT gate, and investigate the time in the cases of allowed errors $10^{-2}, 10^{-3}, 10^{-4}$ and $10^{-5}$ for System E with discretizing $\Omega _3, \Omega _5, \cdots $, where we search the time in only integers as control time, and check if the error of each configuration $\omega \in \Omega _N$ is under a given allowed error which is estimated by Eq.~(\ref{eq:error}).  We also investigate System A to compare the results of System E.

\begin{figure*}[t!]
\begin{minipage}[t]{0.99\textwidth}
\centering
\subfloat[System A]{\includegraphics[width=8.2cm]{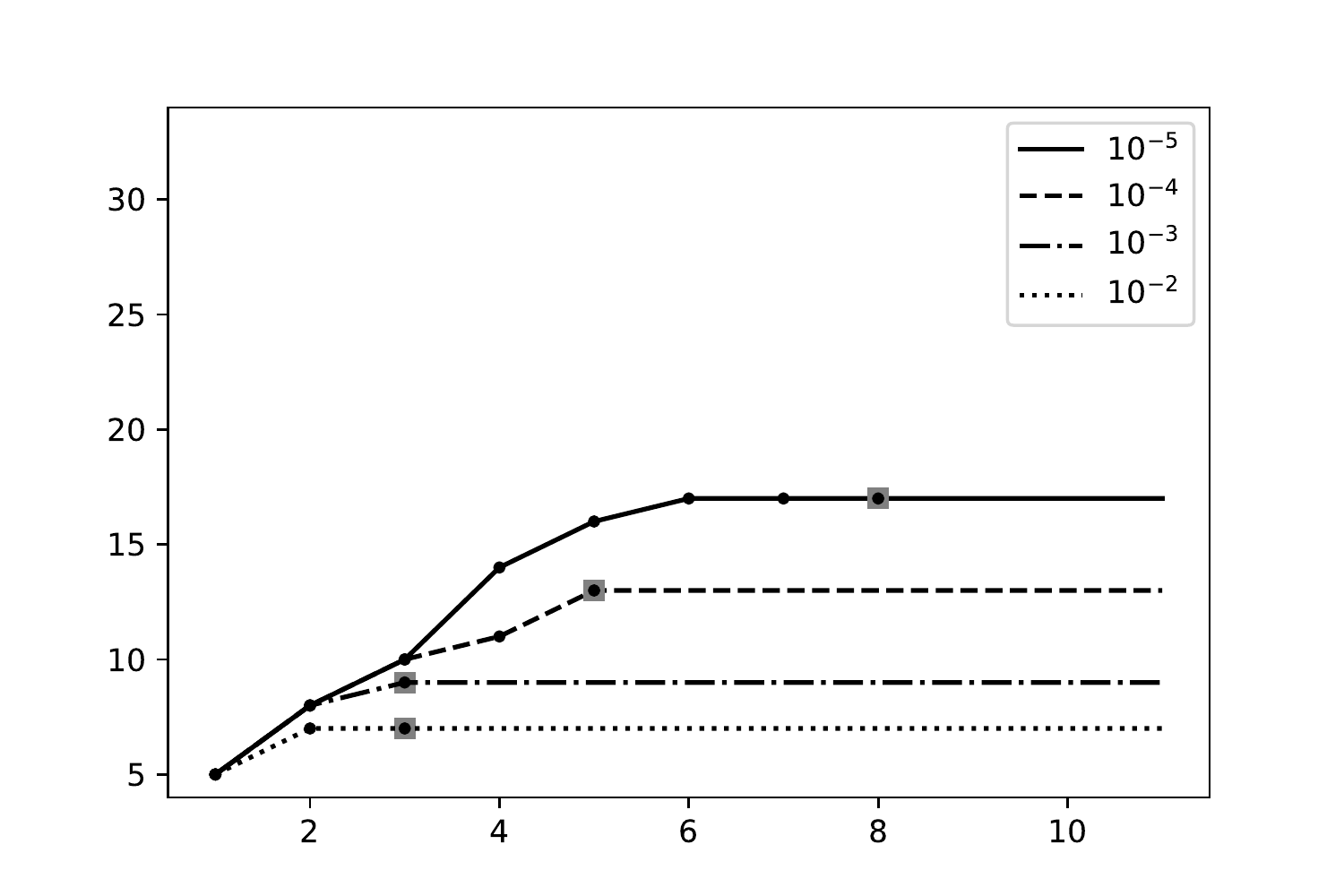}\label{fig:NTA}} \hspace{35pt}
\subfloat[System E]{\includegraphics[width=8.2cm]{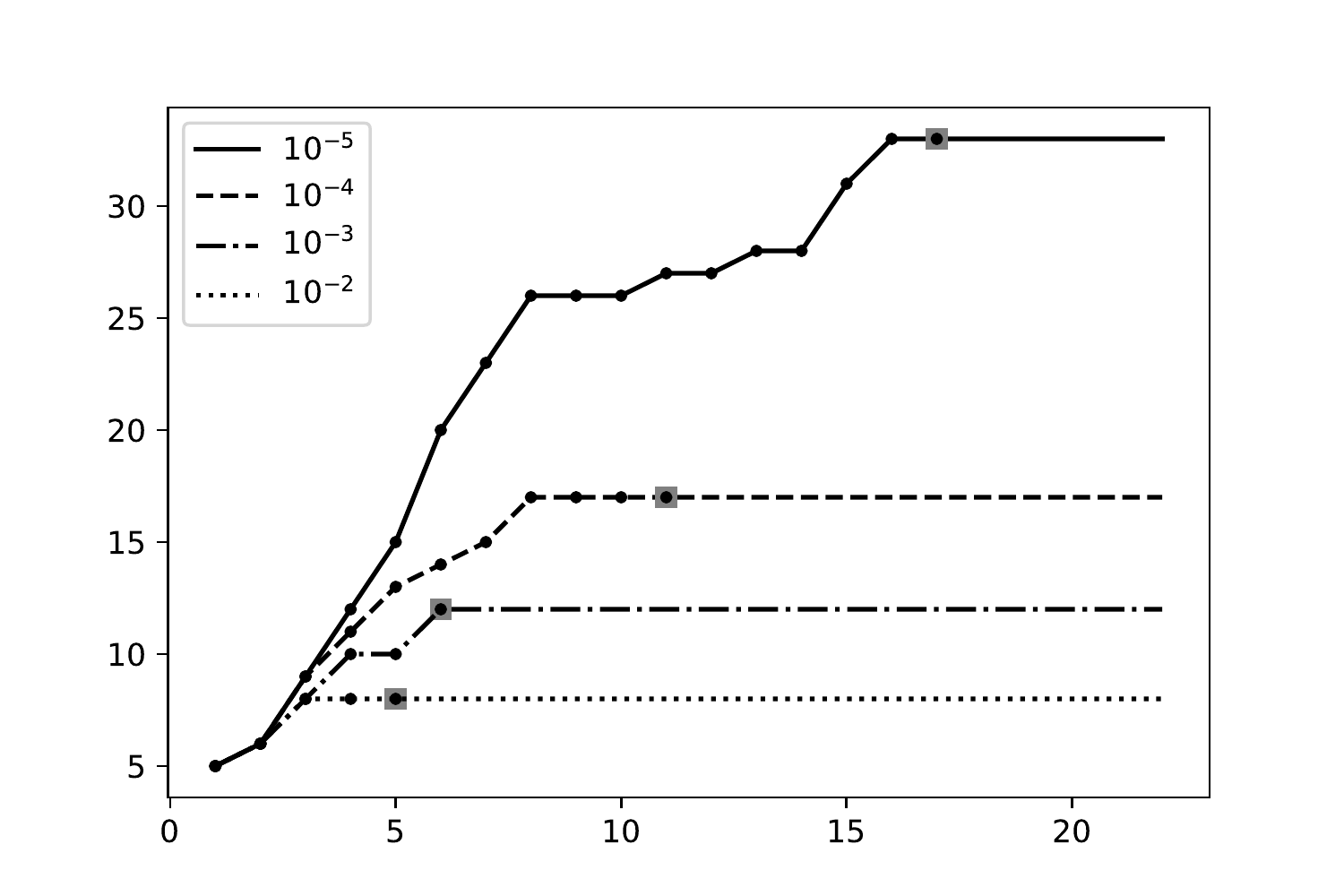}\label{fig:NTE}}
\caption{
	The minimum control time $T_\varepsilon (N)$ (a vertical axis) to let all the $N$ (a horizontal axis) points in $\Omega_N$ be under a given allowed error (a color of lines) where $\Omega _1 := \{1.0 \}$ where we search the time in only integers.	The black dots show the minimum control time $T_\varepsilon (N)$ achieving the CNOT gate within each of allowed errors ($10^{-2}, 10^{-3}, 10^{-4}, 10^{-5}$) for all $\omega \in \Omega _N$, and we can find a robust control pulse to be under a given allowed error for all $\omega \in [1, 2]$ on points with a red square.
}
\label{fig:NT}
\end{minipage}
\begin{minipage}[t]{0.01\textwidth}
	\begin{overpic}[scale=0.1]{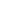}
	\put(-19500,9700){$N$}
	\put(19500,9700){$N$}
	\put(-36000,29600){$T_\varepsilon (N)$}
	\put(2750,29600){$T_\varepsilon (N)$}
\end{overpic}
\end{minipage}
\end{figure*}

	The results of System A and E are shown in Fig.~\ref{fig:NTA} and \ref{fig:NTE}, respectively.	They represent the minimum control time $T_\varepsilon (N)$ (a vertical axis) to let all the $N$ (a horizontal axis) points in $\Omega _N$ be under a given allowed error (a color of lines) where $\Omega _1 := \{ 1.0 \}$.	The black dots show the minimum control time $T_\varepsilon (N)$ achieving the CNOT gate within each of allowed errors ($10^{-2}, 10^{-3}, 10^{-4}, 10^{-5}$) for all $\omega \in \Omega _N$.  We can find a robust control pulse to be under a given allowed error for all $\omega \in [1,2]$ on points with a red square.

System E looks robustly controllable since we can find the robust control pulse for $\omega \in [1,2]$ under $10^{-5}$ error, which is the same level as System A, although this result does not mean the existence of the control pulses achieving an arbitrary small error.   The numerical search for the robust control pulse for System E is more difficult than that for System A since the large $N$ is required to obtain the pulses for a given allowed error to implement the CNOT gate.	This tendency also appears in the cases of implementing another target unitary gate or the single-qubit systems.   For example, this tendency appears when we compare the system with $H_d(\omega) = X + \omega Y$ and the system with $H_c = Z$ with $H_d(\omega) = X + \omega Z$ and $H_c = Z$, where the former system is shown to be robustly controllable but the latter is not.

\section{Summary and Discussions}
\label{sec:Conclusion}

We performed a Lie-algebraic analysis on robust controllability for two-qubit systems, and showed that there exist robustly controllable two-qubit systems by constructing examples (System A, B, C and D).   In the examples we have shown, the control pulse $v(t)$ of the control Hamiltonians is applied only on one of the two-qubits and the robust control is achieved for an unknown parameter in a compact and positive (or negative) continuous set.   We also numerically analyzed the robust controllability of Systems A and found the robust control pulses by using the QuTip control package.  Then we numerically investigated a system whose robust controllability is analytically unclear (System E), and we obtained a robust control pulse achieving around $10^{-4}$ error for all $\omega \in [1, 2]$ as shown in Fig.~\ref{fig:omegaHei1}.   To study the robust controllability of System E within an arbitrarily small error, we investigate the minimum control time for a given $N$-discretized $\omega \in \Omega _N$ by the same numerical approach.	As a result, the robust controllability of under $10^{-5}$ error for all $\omega \in [1, 2]$ on System A and E are numerically shown in Fig.~\ref{fig:NT}.	Thus, we conjecture that System E is also robustly controllable for continuous unknown $\omega$, and any systems which satisfy the conditions \textit{(1)} and \textit{(2)} presented in Sec.~\ref{sec:D} \cite{D:OWR2012}.	It is worth noting that the difficulty of finding the robust control pulses via the QuTip control package may be related with the invertibility of a drift Hamiltonian as we observed this tendency in our numerical results.

\vspace{0.3cm}
\section*{Acknowledgements}

We thank Alexander Pitchford for assisting our numerical calculations.	This work was supported by MEXT Quantum Leap Flagship Program (MEXT Q-LEAP) Grant Number JPMXS0118069605, JSPS KAKENHI (Grant No.15H01677, No.16H01050, No.17H01694, No.18H04286, No.18K13467), and ALPS.

\end{document}